\documentclass{article}%
\usepackage{amsmath}
\usepackage{amsfonts}
\usepackage{amssymb}
\usepackage{graphicx}%
\providecommand{\U}[1]{\protect\rule{.1in}{.1in}}

\begin{document}

\begin{center}
\textbf{AN EFFECTIVE TORSION MODEL FOR EXTRA GAUGE BOSONS AND THEIR MASS
CORRECTIONS}\bigskip

\bigskip J. A. HELAYEL-N\"{E}TO \bigskip

Centro Brasileiro de Pesquisas F\'{\i}sicas,

Rua Dr. Xavier Sigaud 150, Urca,

Rio de Janeiro, Brazil, CEP 22290-180

\bigskip N. PANZA \bigskip

Centro Federal de Educa\c{c}\~{a}o Celso Suckow da Fonseca

Avenida Maracan\~{a}, 229, Maracan\~{a}

Rio de janeiro, Brazil, CEP 20271-110

\bigskip

\textbf{ABSTRACT}

\bigskip
\end{center}

LHC provides a excellent laboratory to probe massive gravitons effects in
scenarios with low scale gravity up to several Tev. Based on this fact, in the
present work we are interested in analyzing the possible constraints on the
free parameters of the gravity Lagrangian in 4 dimensions recently studied in
the reference [1] taking only the torsion effects into account, in order to
generate mass for the graviton. We shall consider a effective fields theory
approach, in which the gravity is excited by the torsion but not the metric.

\bigskip

\bigskip\textbf{I. INTRODUCTION} \bigskip

Beyond questions, Einstein's field theory accounts very well almost all known
macroscopic gravitational phenomena. However, as a quantum theory
(ultra-violet scales) it is not a satisfactory. The extensions of the General
Relativy (GR), considering Lagrangians with powers of the curvature and
torsion, there proved insufficient to solve simultaneously the problems of
renormalizability and unitarity of the theory. In this context, it is
important to analyse the possible effect of propagating torsion in effective
models. In spite of that, as shown in Ref. [1], ever in this case, the
interaction of the particles of the standard model with the propagating
torsion leads to a violation of unitarity, within the first loop quantum
corrections. Thus, for the effects of the dynamic torsion be in conformity
with the phenomenology of the SM and that be important of the effects of low
energy, we must assume that the caractheristc mass of the torsion terms are
much less than the Planck mass [2].

In Cosmology, the observed current accelerated expansion of the Universe might
be signaling a failure of GR at larges distances (infra-red scales), is a
compelling idea that motives the investigation of large distances
modifications of gravity. The GR can be viewed a theory of a massless spin 2
particles, gravitons, but does not excluded the question of whether gravity
could be mediated by a small massive graviton. In this context, in Cosmology
the massive graviton dark matter scenario was condidered as an possible
alternative dark energy model to explaining the present accelerating universe
[3,4]. Furthermore, several peoples agree with the fact that unification
between gravity and electromagnetism have to take into account torsion in four
and high dimensional theories such as Kaluza-klein ones. With all these
arguments it is unquestionable the importance of the torsion field in the
decription of gravity as a fundamental interaction. From the experimental
point of view, the more exciting expectation is the production of a real
massive graviton at LHC from hadronic process and the feasibility of
detectation of quantum gravity effects at the Tev scale.

The aim of the present paper is to propose a unitary and free tachyions model
that generates mass for the graviton, assuming that the masse of graviton
comes only from the excitation of the torsion degree of freedom, in the
so-called Cartan restrict geometry. Specifically, we investigate, in this
context, the model considered in Ref. [1]. We work with the vielbein and the
spin connection as independent fields. This choice is motived because we think
that it is a more fundamental approach to gravitation, since it is based on
the fundamental ideas of the Yang-Mills approach.

This paper is organised as follows: in section II we present a description of
the model, our conventions and obtain the propagators of the corresponding
models. Also, since that Dirac fields couple to the affine connection the
intrinsic spin of fermions acts like a source of torsion, in section III we
study a extension of the our model in order to include the minimal coupling
between torsion and Dirac spinor as well as between torsion and Majorana
spinor. Section IV present discussions and conclusions.

\bigskip

\textbf{II. DESCRIPTION\ OF\ THE\ MODEL}

\bigskip

As is well known, in the space-time in which torsion is present, the metric
and the affine connection can be, in general, considered independent fields.
The last is non-symmetric and the torsion tensor is defined as $T_{\mu\nu}$
$^{\alpha}=\Gamma_{\mu\nu}$ $^{\alpha}-\Gamma_{\nu\mu}$ $^{\alpha}$. In four
dimensions, the torsion tensor has 24 degrees of freedom and can be decomposed
with respect to the Lorentz group into three irreducible tensors, namely: a
pseudo-vector $a^{\lambda}$, a vector $v^{\mu}$ and a rank-three tensor
$q_{\mu\nu\alpha}$, which satisfies the the conditions $q_{\mu\alpha}$
$^{\alpha}=0$ and $\epsilon^{\alpha\beta\mu\nu}q_{\alpha\beta\mu}=0$.
\begin{equation}
T_{\nu\mu\alpha}=q_{\nu\mu\alpha}+\frac{1}{3}\left(  \eta_{\nu\alpha}v_{\mu
}-\eta_{\mu\alpha}v_{\nu}\right)  +\epsilon_{\nu\mu\alpha\lambda}a^{\lambda}
\tag{1}%
\end{equation}

Our purpose is to investigate the role of a propagating torsion \ in the
description massive gravity with explic torsions terms. Within all possible
quadratic terms that we can form with torsion, the independent contributions
turn out to be $T_{\mu\nu}$ $^{a}T^{\mu\nu}$ $_{a}$, $T_{\mu a}$ $^{a}T^{\mu
b}$ $_{b}$, $T_{abc}T^{abc}$. The most general parity-preserving Lagrangian
without higher derivatives is given by [1]:

\begin{equation}%
\mathcal{L}%
_{g}=e\left(  -\frac{1}{\kappa^{2}}R+\chi R^{2}+\beta R_{\mu a}R^{\mu
a}+\gamma R_{\mu a}R^{a\mu}+\xi R_{\mu\nu ab}R^{ab\mu\nu}+kR_{\mu\nu
ab}R^{ab\mu\nu}\right.  \tag{2}%
\end{equation}%
\[
\left.  +\lambda R_{\mu\nu ab}R^{\mu a\nu b}+sT_{\mu\nu}\text{ }^{a}T^{\mu\nu
}\text{ }_{a}+tT_{\mu a}\text{ }^{a}T^{\mu b}\text{ }_{b}+rT_{abc}%
T^{abc}\right)
\]
where $e_{\mu}^{a}$ is the vielbein, $e=\det$ $e_{\mu}^{a}$ and $R_{\mu\nu}$
$^{ab}$ is the field strength associated with the spin connection $\omega
_{\nu}$ $^{ab}$. Moreover, $\chi$, $\beta$, $\gamma$, $\xi$, $k$ and $\lambda$
are arbitrary dimensionless constants and the others parameters have canonical
dimensions given by $\left[  s\right]  =\left[  t\right]  =$ $\left[
r\right]  =1$. Our conventions are [5]:%
\[
R_{\mu\nu}\text{ }^{ab}=\partial_{\mu}\omega_{\nu}\text{ }^{ab}-\partial_{\nu
}\omega_{\mu}\text{ }^{ab}+\omega_{\mu}\text{ }^{a}\text{ }_{c}\omega_{\nu
}\text{ }^{cb}-\omega_{\upsilon}\text{ }^{a}\text{ }_{c}\omega_{\mu}\text{
}^{cb}%
\]%
\begin{equation}
R_{\mu}\text{ }^{a}=e_{b}^{\nu}R_{\mu\nu}\text{ }^{ab} \tag{3}%
\end{equation}%
\[
R=e_{a}^{\mu}e_{b}^{\nu}R_{\mu\nu}\text{ }^{ab}%
\]%
\[
\eta_{ab}=\left(  1,-1,-1,-1\right)
\]
where the Geek indices refer to the world manifold and the Latin ones stand
for the frame indices. Before we proceed, we recall that the spin connection
$\omega_{abc}$ may be expressed as%
\begin{equation}
\omega_{\mu}\text{ }^{ab}=\widetilde{\omega}_{\mu}\text{ }^{ab}+K_{\mu}\text{
}^{ab} \tag{4}%
\end{equation}
which $\widetilde{\omega}_{\mu}$ $^{ab}$ is the Riemannian part of the spin
connection%
\begin{equation}
\widetilde{\omega}_{\mu}\text{ }^{ab}=\frac{1}{2}e_{\mu c}\left(  \Omega
^{cab}+\Omega^{acb}-\Omega^{bac}\right)  \tag{5}%
\end{equation}
where $\Omega_{cba}=e_{c}^{\mu}e_{b}^{\nu}\left(  \partial_{\mu}e_{\nu
a}-\partial_{\nu}e_{\mu a}\right)  $ stands for the rotation coefficients.
$K_{\mu}$ $^{ab}$ are the components of the contorsion tensor
\begin{equation}
K_{\mu}^{ab}=\frac{1}{2}\left(  T_{\mu}\text{ }^{ab}-T^{ab}\text{ }_{\mu
}+T^{b}\text{ }_{\mu a}\right)  \tag{6}%
\end{equation}

As we consider only quadratic terms in the curvature and torsion in the
Lagrangian, by virtue of the Gauss-Bonet theorem,%

\begin{equation}%
{\displaystyle\int}
d^{4}xe\left(  R_{\mu\nu ab}R^{\mu\nu ab}-4R_{\mu a}R^{\mu a}+R^{2}\right)  =0
\tag{7}%
\end{equation}
there is a redundant term among the possibilities for D=4.

Adopting the fact that we are working on a energy scale in which the gravity
is excited by the torsion, but not by the metric, the vielbein can be
expressed as a Kronecker delta, so that $\widetilde{\omega}_{\mu}$ $^{ab}=0$.
Making use (4 ), (5) and (6), the Lagrangian (2 ) can be rewritten as%
\begin{equation}%
\mathcal{L}%
_{g}=-\alpha R+\beta R_{\mu\nu}R^{\mu\nu}+\gamma R_{\mu\nu}R^{\nu\mu}+\xi
R_{\mu\nu\alpha\beta}R^{\mu\nu\alpha\beta}+kR_{\mu\nu\alpha\beta}R^{\alpha
b\mu\nu} \tag{8}%
\end{equation}%
\[
\left.  +\lambda R_{\mu\nu\alpha\beta}R^{\alpha\beta\mu\nu}R^{\mu\alpha
\nu\beta}+sT_{\mu\alpha\beta}T^{\mu\alpha\beta}+tT_{\mu\alpha\beta}%
T^{\alpha\beta\mu}+rT_{\mu\alpha}\text{ }^{\alpha}T^{\mu\beta}\text{ }_{\beta
}\right)
\]

Throughout the paper we shall consider the Cartan restricted geometry, in
which $q_{\nu\mu\alpha}=0$. The physical meaning of this assumtion is that the
tensor component of the torsion is not coupled to the Dirac spinor field, as
discussed in Ref. [2]. Concerning the spin connection $\omega_{\mu}$ $^{ab}$,
one gets the explicit expression%
\begin{equation}
\omega_{\beta\mu\nu}=\frac{1}{3}\left(  \eta_{\beta\nu}v_{\mu}-\eta_{\mu\beta
}v_{\nu}\right)  +\frac{1}{2}\epsilon_{\beta\mu\nu\lambda}a^{\lambda} \tag{9}%
\end{equation}

In order to investigate the spectrum of our model, we split the bilinear piece
of the Lagrangian as:%
\begin{equation}%
\mathcal{L}%
_{g}=\frac{1}{2}\left(  v^{\mu}\text{ }a^{\mu}\right)  O_{\mu\nu}\left(
\begin{array}
[c]{c}%
v^{\nu}\\
a^{\nu}%
\end{array}
\right)  \tag{10}%
\end{equation}
where the Hermitian bilinear form $O_{\mu\nu}$ has the form%
\begin{equation}
O_{\mu\nu}=%
\begin{pmatrix}
\left(  b\square+f\right)  \theta_{\mu\nu}+\left(  \left(  b+c\right)
\square+f\right)  \omega_{\mu\nu} & 0\\
0 & \left(  e\square+g\right)  \theta_{\mu\nu}+\left(  \left(  e+d\right)
\square+g\right)  \omega_{\mu\nu}%
\end{pmatrix}
\tag{11}%
\end{equation}

$\theta_{\mu\nu}=\eta_{\mu\nu}-\dfrac{\partial_{\mu}\partial_{\nu}}{\square}$
being the projector on the transverse vector states and $\omega_{\mu\nu}$ the
projector on the longitudinal vector states. The elements of the matrix
operator (11) are given by:%
\[
b=-\frac{4}{9}\left(  2\beta+4\xi+\lambda\right)
\]%
\[
c=-\frac{2}{3}\left(  \frac{8\beta}{3}+5\gamma+\frac{4\xi}{3}+4k+\frac
{4\lambda}{3}\right)
\]%
\begin{equation}
d=\gamma+2\xi+4k-3\lambda-\beta\tag{12}%
\end{equation}%
\[
e=-\gamma+4\xi-4k+\beta
\]%
\[
f=\frac{1}{3}\left(  4\alpha+4s-2t+6r\right)
\]%
\[
g=-3\left(  \alpha+4s+4t\right)
\]

\bigskip The non-zero propagators in momentum space are listed below,%
\begin{equation}
i\left\langle vv\right\rangle =\dfrac{9/\left[  4\left(  2\beta+4\xi
+\lambda\right)  \right]  }{p^{2}-m_{1}^{2}}\theta_{\mu\nu}+\frac{3/\left[
2\left(  4\beta+4\xi+2\lambda+5\gamma+4k\right)  \right]  }{p^{2}-m_{2}^{2}%
}\omega_{\mu\nu} \tag{13}%
\end{equation}

\bigskip

where%
\[
m_{1}^{2}=\frac{3\left(  t-2s-3r-2\alpha\right)  }{2\left(  2\beta
+4\xi+\lambda\right)  }\text{, }m_{2}^{2}=\frac{t-2s-3r-2\alpha}{4\beta
+4\xi+2\lambda+5\gamma+4k}%
\]

\bigskip%

\begin{equation}
i\left\langle aa\right\rangle =\frac{1/\left(  \gamma-4\xi+4k-\beta\right)
}{p^{2}-m_{3}^{2}}\theta_{\mu\nu}+\frac{1/\left[  3\left(  \lambda
-2\xi\right)  \right]  }{p^{2}-m_{4}^{2}}\omega_{\mu\nu} \tag{14}%
\end{equation}

where
\begin{equation}
m_{3}^{2}=\frac{3\left(  4s+4t+\alpha\right)  }{\left(  \gamma-4\xi
+4k-\beta\right)  }\text{, }m_{4}^{2}=\frac{4s+4t+\alpha}{\lambda-2\xi}
\tag{15}%
\end{equation}

Now, that we have obtained the spectrum of our model, we must search for the
possible choise of the parameters in the Lagrangian (8), such that the massive
gravitons belong to the Tev scale. As in the next section will be interested
only in the pseudo-vector component of the torsion, in the sequel we present
the condition that must be satisfied \ in order to get \ a unitary model
\ with a propagating massive graviton by analysing eq. (15).
\begin{equation}
s=-\frac{\alpha}{4}\text{, \ }\gamma-4\xi+4k-\beta>0\text{, \ }t>0\text{,
\ \ \ \ \ }\lambda-2\xi>0\text{\ \ } \tag{16}%
\end{equation}

This correspond to the following Lagrangean:%
\begin{align}%
\mathcal{L}%
_{g}  &  =-\alpha R+\beta R_{\mu\nu}R^{\mu\nu}+\gamma R_{\mu\nu}R^{\nu\mu}+\xi
R_{\mu\nu\alpha\beta}R^{\mu\nu\alpha\beta}+kR_{\mu\nu\alpha\beta}R^{\alpha
b\mu\nu}\tag{17}\\
&  -\lambda R_{\mu\nu\alpha\beta}R^{\alpha\beta\mu\nu}R^{\mu\alpha\nu\beta
}-\frac{\alpha}{4}T_{\mu\alpha\beta}T^{\mu\alpha\beta}+tT_{\mu\alpha\beta
}T^{\alpha\beta\mu}+rT_{\mu\alpha}\text{ }^{\alpha}T^{\mu\beta}\text{ }%
_{\beta}\nonumber
\end{align}

\bigskip where the masses of propagating modes are:%
\begin{equation}
m_{3}^{2}=\frac{12t}{\gamma-4\xi+4k-\beta}\text{, \ \ \ }m_{4}^{2}=\frac
{4t}{\lambda-2\xi} \tag{18}%
\end{equation}

\bigskip

\textbf{III. COUPLING WITH FERMIONIC MATTER FIELDS} \bigskip

In recent years, the study of gravity with torsion in presence of fermionic
massive matter has received considerable attention for a long time for the
purpose of formulating general relativity as a gauge theory. Futhermore, as it
is well known, in Einstein-Cartan-Sciama-Kibble (ECSK) gravity, the torsion
becomes a dynamical variable related to the spin density of matter [2 ]. In
this section, we shall be concerned with the minimal coupling between fermions
and torsion, assuming that the metric fluctuations are taken to be very week
as compared to the torsion excitations, in the model discussed in section I.
We tackle the issue a 1-loop mass generation mechanism for the fermion,
considered as a Majorana particle. In this case, the interaction between the
fermion and the torsion takes place by the pseudo-vector component $a_{\mu}$.

The requeriment that the Dirac equation in a gravitational field preserve
local Lorentz invariance yields a direct interaction between torsion and
fermions, so that in a Riemann-Cartan space-time, the Dirac Lagrangian for a
massive fermion has the form%

\begin{equation}%
\mathcal{L}%
_{D}=e\left(  ie_{a}^{\mu}\overline{\psi}\gamma^{a}D_{\mu}\psi-m_{0}%
\overline{\psi}\psi\right)  \tag{19}%
\end{equation}
where the covariant derivative of the spinor field are given by%
\begin{equation}
D_{\mu}=\partial_{\mu}+\frac{g}{8}\omega_{\mu}^{ab}[\gamma_{a},\gamma_{b}]
\tag{20}%
\end{equation}
and $g$ is a dimensionless coupling constant, $\gamma^{a}=\gamma^{\mu}e_{\mu
}^{a}$ with $\gamma^{\mu}$ denoting Dirac matrices and $m_{0}$ is the mass of
the Dirac field.

In order to rewrite the Dirac Lagrangian in terms of the vector and
pseudo-vector components of the torsion, we make use of the identity
\begin{equation}
\gamma^{a}\gamma^{b}\gamma^{c}=\gamma^{a}\eta^{bc}+\gamma^{c}\eta^{ab}%
-\gamma^{b}\eta^{ab}+i\epsilon^{abcd}\gamma_{5}\gamma_{d} \tag{21}%
\end{equation}
for obtain, after some algebra%
\begin{equation}%
\mathcal{L}%
_{D}=i\overline{\psi}\gamma^{\mu}\partial_{\mu}\psi-m_{0}\overline{\psi}%
\psi-i\frac{g}{2}\overline{\psi}v^{\alpha}\gamma_{\alpha}\psi+\frac{3g}%
{4}\overline{\psi}\gamma_{5}a_{\lambda}\gamma^{\lambda}\psi\tag{22}%
\end{equation}

Now, we can extract the Feynman rules for the theory. The propagator for the
pseudo-vector is given by eq. (12)%
\begin{equation}
i\Delta_{0}\left(  p\right)  =\frac{A}{p^{2}-m_{4}^{2}}\eta_{\mu\nu}+\left(
\frac{B}{p^{2}-m_{3}^{2}}-\frac{A}{p^{2}-m_{4}^{2}}\right)  \frac{p_{\mu
}p_{\nu}}{p^{2}} \tag{23}%
\end{equation}

where $A=1/\left(  \gamma-4\xi+4k-\beta\right)  >0$ and $B=1/\left[  3\left(
\lambda-2\xi\right)  \right]  >0$, and for the spinor field is given by%
\begin{equation}
iS_{0}\left(  p\right)  =\frac{1}{p\!\!\!/-m_{0}} \tag{24}%
\end{equation}

Futhermore, the fermion-torsion vertex is given by%
\begin{equation}
V_{\lambda}=i\frac{3g}{4}\gamma_{5}\gamma_{\lambda} \tag{25}%
\end{equation}

From these results, we can calculate the self-energy corrections for the bare
propagators and discuss the mechanism of mass generation for fermion fields.
For the fermion self-energy graph, we find:%
\[
-i\Sigma\left(  p\right)  =\left(  i\right)  ^{4}\left(  \frac{3g}{4}\right)
^{2}\int\frac{d^{4}q}{\left(  2\pi\right)  ^{4}}\gamma_{5}\gamma^{\mu}%
\frac{p\!\!\!/-q\!\!\!/+m_{0}}{\left(  p-q\right)  ^{2}-m_{0}^{2}}\gamma
_{5}\gamma^{\nu}\left[  -\frac{A}{p^{2}-m_{4}^{2}}\eta_{\mu\nu}+\right.
\]%
\begin{equation}
\left.  \left(  \frac{A}{p^{2}-m_{4}^{2}}-\frac{B}{p^{2}-m_{3}^{2}}\right)
\frac{q_{\mu}q_{\nu}}{q^{2}}\right]  \tag{26}%
\end{equation}

\bigskip By a direct calculation, we obtain%
\begin{equation}
\Sigma\left(  p\right)  =\frac{9Am_{0}^{2}}{32\pi^{2}m_{4}^{2}}\ln\Lambda
^{2}g^{2}\mu^{-\frac{1}{2\ln\Lambda^{2}}}\left(  p\!\!\!/-m_{0}\right)
+\text{finite} \tag{27}%
\end{equation}
where $\mu$ is a arbitrary mass parameter and $\Lambda$ is a cutoff. The
finite part of the correction to the propagator is not very important in the
following discussion, and therefore is omitted.

Making use of the unrenormalized proper self-energy part of equation (27),
$\widetilde{\Sigma}\left(  p\right)  $, we can express the observed mass $m$
in terms of the bare mass $m_{0}$, coupling constant $g$ and cutoff $\Lambda$.%

\begin{equation}
m-m_{0}=\widetilde{\Sigma}\left(  p\right)  \left\vert _{p\!\!\!/=m}\right.
\tag{28}%
\end{equation}
or explicitly%

\begin{equation}
m-m_{0}=\left(  m-m_{0}\right)  \frac{9Am_{0}^{2}}{32\pi^{2}m_{4}^{2}%
}g_{\text{eff}}^{2} \tag{29}%
\end{equation}

\bigskip where $g_{\text{eff}}=g\left(  \dfrac{\ln\Lambda^{2}}{\mu^{1/\left(
2\ln\Lambda^{2}\right)  }}\right)  ^{1/2}$is a effective coupling constant.

Our explicit calculations indicate that the gravitational coupling of the
fermion to the pseudo-vector associated to torsion degrees of freedom is not
able to dynamically induce mass. We check that, by setting $m_{0}=0$, $m$
turns out to vanish. So dynamical mass generation does not take place. so, the
only possible effect of the coupling to $a^{\mu}$ is a mass shift upon
radiative corrections.

\ \ \ \ \ We then adopt the following scenario. The tree-level mass, $m_{0}$,
of the fermion must have been generated by means of the spontaneous breakdown
of some (local) internal symmetry through a higgs field. We suppose some extra
massive higgs of the MSSM [ ] acquired a vacuum expectation value and endowed
the fermion with mass $m_{0}\sim1$ Tev. Then, by invoking naturalness [ ], we
expect a shift of, at most, the same order. This means that the gravitational
coupling shoud produce a shift of the order of $1$ Tev. With that physical
assumption, our 1-loop computations yield the constraint below for the
effective gravitational coupling, $g_{\text{eff}}$:%
\begin{equation}
g_{\text{eff}}\leq\frac{4\pi}{3}\sqrt{\frac{2}{A}} \tag{30}%
\end{equation}

However, here a non-trivial fact occurs: the shifted mass, $m$, happens to
dropout of our gap-like equation. Therefore, the most we can infer is a bound
on $g$ set up by $m_{0}^{2}$. it is very unexpected that, at the very end,
only $m_{0}^{2}$ remains in the would-be gap equation. The coupling to thje
vector boson, a gauge-type interaction, is the responsible for this result.
Actually, it is not generally proved, but only 4-fermion interactions give
rise to dynamical mass generation.

\bigskip

III. CONCLUDING COMMENTS \bigskip

\ \ \ \ \ We have taken in our approach the viewpoint that torsion is the only
carrier of the gravity degrres of freedom. Fluctuations of the vielbein could
also be taken into account that yield mixing terms between the emerging
effective vector bosons [9]. This would allow us to build up complex
combinations of real gauge bosons to include extra charged mediators in our
proposal. For that, the parameters should be such that they allow a degeneracy
in mass between the charged gauge vectors that are the anti-particle of one
another. We shall be reporting on this development in a forthcoming paper.

On the other hand, we expect, on the basis of our considerations of the
phenomenology of quark-gluon plasmas at ALICE/LHC and RHIC, that the effects
of the mixings shoud be rather small with respect to the contributions that
torsion alone yields. This is so because the average energy density is very
high [10] and fluctuations of the vielbein/metric become less relevant, for we
expect the density of gravitons to be high. But, anyway, we believe, as
already anticipated, that, even if the torsion fluctuations should dominate
over the metric excitation modes, we should exploit in more details the
mixings, to understand how the torsion fluctuations could eventually
strengthen the metric fluctuations. We keep in mind to go further and to
pursue an investigation on the vielbein-spin conection mixings in terms of the
vector bosons that may be produced as effective fluctuations and to undestand
possible effects of the mixings at the Tev scale.

\bigskip

\textbf{References} \bigskip

[1] C. A. Hernaski, A. A. Vargas-Paredes, J. A. Helay\"{e}l-Neto, Phys. Rev.
D80: 124012, 2009.

[2] G. de Berredo-Peixoto, J.A. Helayel-Neto, I.L. Shapiro, JHEP 0002 (2000)
003 .

[3] S. Capozziello, R. Cianci, C. Stornaiolo,Class.Quant.Grav.24:6417-6430,2007.

[4] Hyung Won Lee, Kyoung Yee Kim, Yun Soo Myung, General Relativity\ and
Quantum Cosmology (gr-qc); Cosmology and Extragalactic
Astrophysics\ (astro-ph.CO), 2001.

[5] V. de Sabbata and M. Gasperini, Introduction to Gravitation (World
Scientific, Singapore,1985).

[6] Eric A. Bergshoeff, J.J. Fern%
\'{}%
andez-Melgarejo, Jan Rosseel, Paul K. Townsend, JHEP (2012).

[7] J. L. Boldo, J. A. Helayel-Neto, N. Panza , Class.Quantum Grav. 19 (2002)
2201 .

[8] Nikodem J. Pop lawski, Gen.Relativ.Gravit.44:491-499,2012.

[9] R. Foot and X.-G. He, Phys. Lett. B267 (1991) 509; F. Del Aguila, G. D.
Coughlam and M. Quiros, Nucl. Phys. B307, (1988) 633; B. Holdom, Phys.
Lett.\ B259 (1991) 329; M.Gasperini, Phys. Lett. B263 (1991) 267.

[10] E. Scapparone, "Latest results from ALICE", arXiv:1111.2685 [nucl-ex], on
behalf of the ALICE Collaboration.

\end{document}